\font\fourteenbf=cmbx12 at 14pt
\font\fourteenrm=cmr12 at 14pt
\font\fourteenit=cmti12 at 14pt
\font\twelvebf=cmbx12
\font\twelverm=cmr12
\font\twelveit=cmti12
\font\twelvesl=cmsl12
\font\tenbf=cmbx10
\font\tenrm=cmr10
\font\tenit=cmti10
\font\tenex=cmex10
\font\tensl=cmsl10
\font\ninebf=cmbx9
\font\ninerm=cmr9
\font\nineit=cmti9
\font\nineex=cmex9
\font\ninesl=cmsl9
\font\eightbf=cmbx8
\font\eightrm=cmr8
\font\eightit=cmti8
\font\eightex=cmex8
\font\eightsl=cmsl8
\font\sevenbf=cmbx7
\font\sevenrm=cmr7
\font\sevenit=cmti7
\font\sevenex=cmex7

\def\fourteen{\let\rm=\fourteenrm \let\bf=\fourteenbf \let\it=\fourteenit
            \let\sl=\fourteensl}
\def\twelve{\let\rm=\twelverm \let\bf=\twelvebf \let\it=\twelveit 
            \let\sl=\twelvesl}
\def\ten{\let\rm=\tenrm \let\bf=\tenbf \let\it=\tenit
         \let\ex=\tenex \let\sl=\tensl}
\def\nine{\let\rm=\ninerm \let\bf=\ninebf \let\it=\nineit
          \let\ex=\nineex \let\sl=\ninesl}
\def\eight{\let\rm=\eightrm \let\bf=\eightbf \let\it=\eightit
           \let\ex=\eightex \let\sl=\eightsl}
\def\seven{\let\rm=\sevenrm \let\bf=\sevenbf \let\it=\sevenit
           \let\ex=\sevenex } 

\hsize=6.0truein
\vsize=8.6truein

\baselineskip=15pt
\parindent=15pt
\twelve

\newif\ifdebug 
\newif\ifbacketedref 
\newif\ifpareqref 
\newif\ifpareqno 
\newif\ifeqnoperchapter 
\def\chapterfont{\twelvebf}

\def\titlefont{\twelvebf}
\def\centerline#1{\line{\hss{#1}\hss}}%
\def\title#1{{\titlefont\centerline{#1}\vskip 5mm}}
\def\authors#1{\centerline{#1} }
\def\DURHAM{{\parskip 2mm\tenit\baselineskip=13pt
    \centerline{Department of Mathematical Sciences}
    \centerline{University of Durham, Durham DH1 3LE, England}
}}

\def\Email#1{{\parskip 2mm\tenit\baselineskip=13pt\centerline{E-Mail: #1}}}

\def\abstract#1{{\tenrm\baselineskip=13pt
\parindent=0pt\vskip 1cm
\centerline{ABSTRACT}
\vbox{}
\vbox{\leftskip=12mm\rightskip=12mm#1}
}}
\newtoks\date 
\def\monthname{\relax\ifcase\month 0/\or January\or February\or
    March\or April\or May\or June\or July\or August\or September\or
    October\or November\or December\else\number\month/\fi}
\date={\monthname\ \number\day, \number\year}
\def\pubnum#1{{\twelverm\obeylines\everypar{\hfill} 
\rm DTP-#1

\the\date}
\vskip 5mm}
\newcount\chapternumber \chapternumber=0 
\newcount\sectionnumber \sectionnumber=0
\def\chapterreset{%
   \global\advance\chapternumber by 1%
   \ifeqnoperchapter \global\equanumber=0 \fi
   \sectionnumber=0 
}
\def\chapter#1{
{ \parindent=0pt \vglue 0.5cm
  \baselineskip=13pt
  \chapterreset
  {\chapterfont \number\chapternumber .\ #1 \hfill}
  \vglue 0.2cm
}}

\newcount\equanumber         \equanumber=0
\def\eqname#1{
  \def\Ename{\string#1}
  \relax 
  \ifnum\the\equanumber<0
    \xdef\LEQno{{\tenrm(\number-\equanumber)}}\global\advance\equanumber by -1
  \else \global\advance\equanumber by 1
    \ifdebug
       \xdef\LEQno{\Ename} 
    \else
      \ifeqnoperchapter
        \xdef\LEQno{\number\chapternumber .\number\equanumber} 
      \else
        \xdef\LEQno{\number\equanumber} 
      \fi
    \fi
  \fi
  \ifpareqref
    \xdef#1{{\rm(\LEQno)}\ }
  \else
    \xdef#1{{\rm Eq.\LEQno}\ }
 \fi
  \ifpareqno {\tenrm(\LEQno)} \else {\tenrm\LEQno} \fi 
}

\def\eqn#1{\eqno\eqname{#1}}

\newif\ifbacketedref
\newif\ifreferenceopen       \newwrite\referencewrite
\newdimen\referenceminspace  \referenceminspace=25pc
\newdimen\refindent          \refindent=30pt
\newcount\referencecount     \referencecount=0
\def\reffile{rf}
\immediate\openout\referencewrite=\reffile
\def\refout{\par \penalty-400 \vskip\chapterskip 
 \immediate\closeout\referencewrite
   \referenceopenfalse
   \vglue 0.5cm
   \baselineskip=13pt
   {\chapterfont References\hfil}
   \vglue 0.2cm
   \baselineskip=15pt
   \input \reffile
   }
\def\Textindent#1{\noindent\llap{#1\enspace}\ignorespaces}
\def\refitem#1#2{\par \hangafter=0 \hangindent=\refindent \Textindent{#1} #2}
\def\refnum#1{\global\advance\referencecount by 1 \def\Rname{\string#1}
\ifdebug\xdef#1{\Rname}\else\xdef#1{\the\referencecount}\fi
}
\def\refmark#1{\hbox{\raise1ex\hbox{{\eightrm%
   \ifbacketedref[#1]\else#1\fi}}}}

\def\Ref#1#2{\refnum#1\refmark{#1}%
  \immediate\write\referencewrite{\noexpand\refitem{#1.}{#2}}%
}
\def\REF#1#2{\refnum#1%
 \immediate\write\referencewrite{%
 \noexpand\refitem{#1.}{#2}}%
}

\newskip\headskip 	\headskip=8pt plus 3pt minus 3pt
\newskip\chapterskip    \chapterskip=\bigskipamount
\newskip\sectionskip     \sectionskip=\medskipamount
\def\ack{\par\penalty-100\medskip 
    \line{\twelverm\hfil ACKNOWLEDGEMENTS\hfil}\nobreak\vskip\headskip }

\newcount\Tableno     \Tableno=0
\def\figureinc{%
   \global\advance\figureno by 1%
}
\def\Table#1#2#3{\vskip 10mm%
{#3}
\global\advance\Tableno by 1%
\xdef#1{{\rm\number\Tableno}\ }%
\vskip 2mm
\centerline{Table \number\Tableno\ : #2}
\vskip 10mm
}
\def\NextTableNo{{\advance\Tableno by 1 \number\Tableno}}
%
\newcount\figureno     \figureno=0
\newdimen\figdim       \figdim=70mm
\def\figureinc{%
   \global\advance\figureno by 1%
}
\def\figcaption#1#2#3{\hbox to #2{\hss{\vbox{\hsize=#2 \parindent=0pt 
        {\bf Figure \number\figureno#3 :\ }#1}}\hss}
}

\def\OneFig#1#2{\vskip 5mm\figdim=70mm
\centerline{\figureinc
  \vbox{\epsfxsize=6cm\epsfysize=6cm\epsfbox{#1}\vskip 5mm
        \figcaption{#2}{\figdim}{}}
  }\vskip 5mm
}
\def\TwoFigs#1#2#3#4{\vskip 5mm\figdim=70mm
 \hbox to \hsize {
  \vbox {\figureinc\epsfxsize=7cm\epsfysize=7cm\epsfbox{#1}\vskip 5mm
         \figcaption{#2}{\figdim}{}
  }\hfill
  \vbox {\figureinc\epsfxsize=7cm\epsfysize=7cm\epsfbox{#3}\vskip 5mm
         \figcaption{#4}{\figdim}{}}}
 \vskip 5mm
}
\def\TwoFigsAB#1#2#3#4{\vskip 5mm\figdim=70mm
 \hbox to \hsize {\figureinc
  \vbox {\epsfxsize=7cm\epsfysize=7cm\epsfbox{#1}\vskip 5mm
         \figcaption{#2}{\figdim}{.a}
  }\hfill
  \vbox {\epsfxsize=7cm\epsfysize=7cm\epsfbox{#3}\vskip 5mm
         \figcaption{#4}{\figdim}{.b}}}
 \vskip 5mm
}
\def\FourFigsAD#1#2#3#4#5#6#7#8{\vskip 5mm\figdim=70mm\figureinc
 \hbox to \hsize {
  \vbox {\epsfxsize=7cm\epsfysize=7cm\epsfbox{#1}\vskip 5mm
         \figcaption{#2}{\figdim}{.a}
  }\hfill
  \vbox {\epsfxsize=7cm\epsfysize=7cm\epsfbox{#3}\vskip 5mm
         \figcaption{#4}{\figdim}{.b}}}
 \vskip 5mm
 \hbox to \hsize {
  \vbox {\epsfxsize=7cm\epsfysize=7cm\epsfbox{#5}\vskip 5mm
         \figcaption{#6}{\figdim}{.c}
  }\hfill
  \vbox {\epsfxsize=7cm\epsfysize=7cm\epsfbox{#7}\vskip 5mm
         \figcaption{#8}{\figdim}{.d}}}
 \vskip 5mm
}
\def\cmod#1{ \vert #1 \vert ^2 }
\def\mod#1{ \vert #1 \vert }
\def\ie{\hbox{\it i.e.\ }}

\def\Sinh{\hbox{sinh}}

\def\C {{\rlap{\kern 1.0mm \vrule height 7pt depth 0pt} \rm C}}
\def\R {{\rlap{\kern 0.1mm \vrule height 7pt depth 0pt} \rm R}}
\def\U {{\rlap{\kern 1.2mm \vrule height 7pt depth 0pt} \rm 1}}

\debugfalse
\backetedreftrue
\eqnoperchaptertrue
\pareqreftrue 
\pareqnotrue
\eqnoperchaptertrue

\twelve
\rm
 
\input epsf


\def\rLak{M. Daniel, K. Porsezian and M. Lakshmanan - 
{\it J. Math. Phys.} {\bf 35}  12 (1994)}

\def\rpra{see eg. A.P. Malozemoff and J.C. Slonczewski, 
Magnetic Domain Walls in Bubble Materials (Academic Press, New York) (1979)}

\def\rnicos{N. Papanicolaou and T.N. Tomaras, {\it Nucl. Phys.} {\bf B 360}
425 (1991), N. Papanicolaou, {\it Physica} {\bf D 74} 107 (1994) }
\def\rNP{N. Papanicolaou and W.J. Zakrzewski, {\it Physica} {\bf D 80} 225-245
(1995)}
\def\rNPZ{N. Papanicolaou and W.J. Zakrzewski, {\it Phys. Lett.} {\bf A 210} 328-336
(1996)}
\def\rgolden{T.H. O'Dell, Ferromagnetodynamics, the dynamics
of magnetic bubbles, domains and domain walls (Wiley, New York) (1981)}
\def\rStraTam{G.N. Stratopoulos and T.N. Tomaras - Vortex Pairs in charged
fluids - Crete preprint 96-10 (hep-th/9601172)}

\pubnum{96/47}

\title{Localized Solutions in a 2 Dimensional Landau-Lifshitz  Model}
\authors{B. Piette,}
\authors{and}
\authors{W.J. Zakrzewski}
\DURHAM
\Email{B.M.A.G.Piette@uk.ac.durham\quad W.J.Zakrzewski@uk.ac.durham}
 
\abstract{We demonstrate the existence of stable time dependent solutions of
 the Landau-Lifshitz 
model with a constant external magnetic field. 
We find such solutions in all topological sectors, including N=0.
We discuss some of their
properties.
}

\chapter{Introduction.}

The dynamics of magnetic bubbles is an issue of practical interest
\Ref\Rpra{\rpra}. A few years ago 
 Papanicolaou and Tomaras\Ref\Rnicos{\rnicos}
showed how to construct  unambiguous conservation laws for 
systems described by the Landau-Lifshitz equation
and pointed out the similarity
of the gross features of the bubble dynamics to
those of the familar Hall effect. This has stimulated
more theoretical and numerical research. In particular 
in \REF\RNP{\rNP}[\RNP] an attempt was made to describe
the dynamics of two bubbles. The model used in [\RNP] was (2+1) dimensional
and to stabilise the bubbles an additional ``Skyrme-like" term
was added to the more familar exchange, anisotropy and ``magnostatic"
terms. It was observed that the bubbles rotate around each other
and that the dynamics of a system of one bubble and one antibubble exhibits
the familar Hall motion.

\REF\RNPZ{\rNPZ}
The system in ref. [\RNP] was modified in [\RNPZ] to include an external 
magnetic field. This allowed the study of the so-called ``golden
rule" of bubble dynamics\Ref\Rgolden{\rgolden}. The introduction 
of the external magnetic field has suggested that we should 
have a serious look 
at the solution of the Landau Lifshitz equation with an (external)
magnetic field. Hence in this paper we look at the case
when this magnetic field is produced by the anisotropy of the system;
thus we look at solutions of the Landau-Lifshitz equation
for the anisotropic Heisenberg model here thought of as describing
a magnetisation field.

So, the equation we want to study is
$$
\partial_{t}\vec\phi = \vec\phi \wedge \bigl[ \nabla^2 \vec\phi + 
A (\vec\phi.\vec n) \vec n\bigr] 
\eqn\eEqPhi
$$
where $\phi = (\phi_1, \phi_2, \phi_3)$ is a unit vector describing 
the orientation of the magnetization,
and $\vec n = (0,0,1)$ is the  vector of the external 
magnetic field (or, put in other words, the direction of anisotropy).
We shall assume, in what follows, that the magnetization is defined over the
two-dimensional plane $\vec x = (x,y)$. 

There are two classes of interesting models describing such systems 
and they differ
by the type of boundary condition that the magnetisation
 takes at infinity. There the field $\vec\phi$
can be required to be perpendicular to $\vec n$ 
(this defines the so called easy plane model) or we can require that it is 
parallel to $\vec n$ (this is the easy-axis model). In what follows we will be interested in the ``easy-axis" model and, for this reason,
 we will consider the following
expression for the total energy of the system  
$$
E = {1 \over 8 \pi} \int dx dy 
      \Bigl[ (\partial_{x} \vec\phi . \partial_{x}\vec\phi) + 
       (\partial_{y} \vec\phi . \partial_{y}\vec\phi) 
       + A (1-\phi_3^2)\Bigr],
\eqn\eEnPhi
$$
where $A$ is a positive constant.

As one can be easily shown the time evolution \eEqPhi\ preserves the
energy \eEnPhi. Another conserved quantity is given by the topological charge
$$
	Q = {1 \over 4 \pi} \int dx dy \Bigl[ \vec\phi . 
             (\partial_{x}\vec\phi \wedge \partial_{y}\vec\phi)\Bigr].
\eqn\eQPhi
$$

The fact that the magnetization is required to take
a single value at infinity means that the two dimensional plane can be 
thought of as being
compactified to a sphere and the topological charge \eQPhi 
which is normalised to take integer values, simply
describes the number of times the field wraps itself around the sphere.

Instead of using the normalized vector $\vec\phi$ to describe the 
magnetization it is convenient to perform 
the stereographic projection of the sphere onto the 
complex plane and use the complex field  
$$
w = {\phi_1 + i \phi_2 \over 1 + \phi_3 }.
$$
 In this formulation the energy functional \eEnPhi\ takes the form 
$$
E = {1 \over 2 \pi} \int dx dy \Bigl[ { \cmod{w_{x}} + \cmod{w_{y}}
                       \over (1 + \cmod{w})^2}
       + { A \cmod{w} \over (1 + \cmod{w})^2 }\Bigr] 
\eqn\eEnW
$$
and the topological density is given by
$$
Q = {1 \over 2 \pi} \int dx dy \Bigl[ { w_{y}^*\ w_{x} - w_{x}^*\ w_{y}
      \over (1 + \cmod{w})^2}
      \Bigr].
\eqn\eQW
$$
Finally, the equation of motion \eEqPhi becomes
$$
i w_t + w_{xx} + w_{yy} - { 2 w^* (w_x^2 + w_y^2) \over 1 + \cmod{w} }
- A w {1 - \cmod{w} \over 1 + \cmod{w}} = 0.
\eqn\eEqW
$$
When the external magnetic field is switched off, $A = 0$,  and the static 
solutions of \eEqW\ are given by holomorphic functions 
of $z=x+iy$, namely, $w = w(x+iy)$.

\chapter{Non Topological Periodic Solutions}
\REF\RLak{\rLak}
In [\RLak] a non static axially symmetric solution of the Landau-Lifshitz 
equation without an external field ($A=0$) was presented and discussed.
This solution was shown to represent a wave-like field
configuration whose energy density is in the form of rings
propagating radially out towards infinity. When a term describing 
the interaction 
with a constant  external magnetic field is added to \eEqPhi\ the 
situation is quite different.
 To see what happens in this case, we have, first of all, taken 
$w = 1/\Sinh(r^2)$ as our initial 
condition and  integrated \eEqW numerically. 

We were very suprised  to see
that, at  first, the ring expended a little and then settled down to a 
configuration periodic in time (after radiating out small rings of energy 
which propagated towards infinity).

A careful analysis of the final field as seen in  our numerical simulation 
indicated that it was of the form 
$$
w = e^{-i \omega t} f(r),
\eqn\eEnRealW
$$ 
where $f(r)$ is a real function.
Inserting the ansatz \eEnRealW\ into \eEqW\ leads to the following 
equation for $f$ 
$$
f_{rr} + {f_r\over r} - {2 f f_r^2 \over 1 + f^2}
- A f { (1 - f^2) \over (1 + f^2)} + \omega f = 0,
\eqn\eEqf
$$ 
and the energy density \eEnW\ becomes
$$
E = \int r dr \Bigl[ {f_r^2 \over (1 + f^2)^2} + {A f^2 \over (1 + f^2)^2} 
              \Bigr].
\eqn\eEnf 
$$
However, the equation \eEqf\ is quite difficult to solve analytically but,
 on the other
hand, it can easily be solved numerically using the shooting method or 
an over-relaxation technique. Before doing so, we notice that 
solutions of \eEqf\ minimize the functional 
$$
D = \int r dr \Bigl[ {f_r^2 \over (1 + f^2)^2} + {A f^2 \over (1 + f^2)^2} -
 { \omega f^2 \over 1 + f^2} \Bigr].
\eqn\eDensf 
$$
From this observation  we note that, 
for any configuration $f$, performing the change of 
variable $r \rightarrow \lambda r$ leaves the first term in $D$ unchanged 
but the last two terms get multiplied by $\lambda^2$. This means than 
unless the last two terms add up to zero we can always
choose $\lambda$ so that the energy of the configuration decreases under that
transformation. This implies that for the non trivial solutions of \eEqf\ 
the last two terms in \eDensf\ must cancel out exactly:
$$
\int r dr {A f^2 \over (1 + f^2)^2} = \int r dr { \omega f^2 \over 1 + f^2}.
\eqn\efCond 
$$   
As for any non-vanishing $f$ we have 
$$
{f^2 \over (1 + f^2)^2} < {f^2 \over 1 + f^2}.
$$
we can directly conclude that for any solution of \eEqf\ we must have
$$
A > \omega.
$$ 
To integrate \eEqf\ we start by analyzing the asymptotic behaviour of
$f$. At infinity $f$ must  vanish to ensure that the total energy
\eEnW\ is finite. At the origin, $f$ must be of the form 
$f = K + Br^2 +O(r^3)$ where $K$ is a constant depending on $A$ and $\omega$,
and where $B = {1 \over 4} \omega K + A K { 1-K^2 \over 1+K^2}$.

As we can always set $A$ to $1$ by redefining $r$ as mentioned above, there is only 
one parameter in the equation which we choose to be $\omega$. 
We have solved \eEqf\ using the shooting method for different values of 
$\omega$ in the range
$0 < \omega < 1$ and have checked that \efCond\ is satisfied for each solution.
In Fig 1 we present the profiles for $\phi_3$ as well as the 
energy density profiles for $\omega = 0.25, 0.5$ and $0.75$. 
We note that when $\omega$ is small the energy density takes the form of 
a ring with a radius
which gets larger as the value of $\omega$ decreases.
We have also found numerically that the total energy of any solution
is approximately given by the relation:
$E = A/\omega$. We thus see that as the frequency increases the energy of 
the solution decreases while, at the same time, the solution becomes more 
localized. Note that all our solutions correspond to field 
configurations where $\phi_3$ goes from a fixed value at the origin to the
vacuum $(\phi_3 = 1)$ at infinity. Moreover, the field  precesses in the
$\phi_1,\phi_2$ plane with the (constant) frequency $\omega$.

\TwoFigsAB{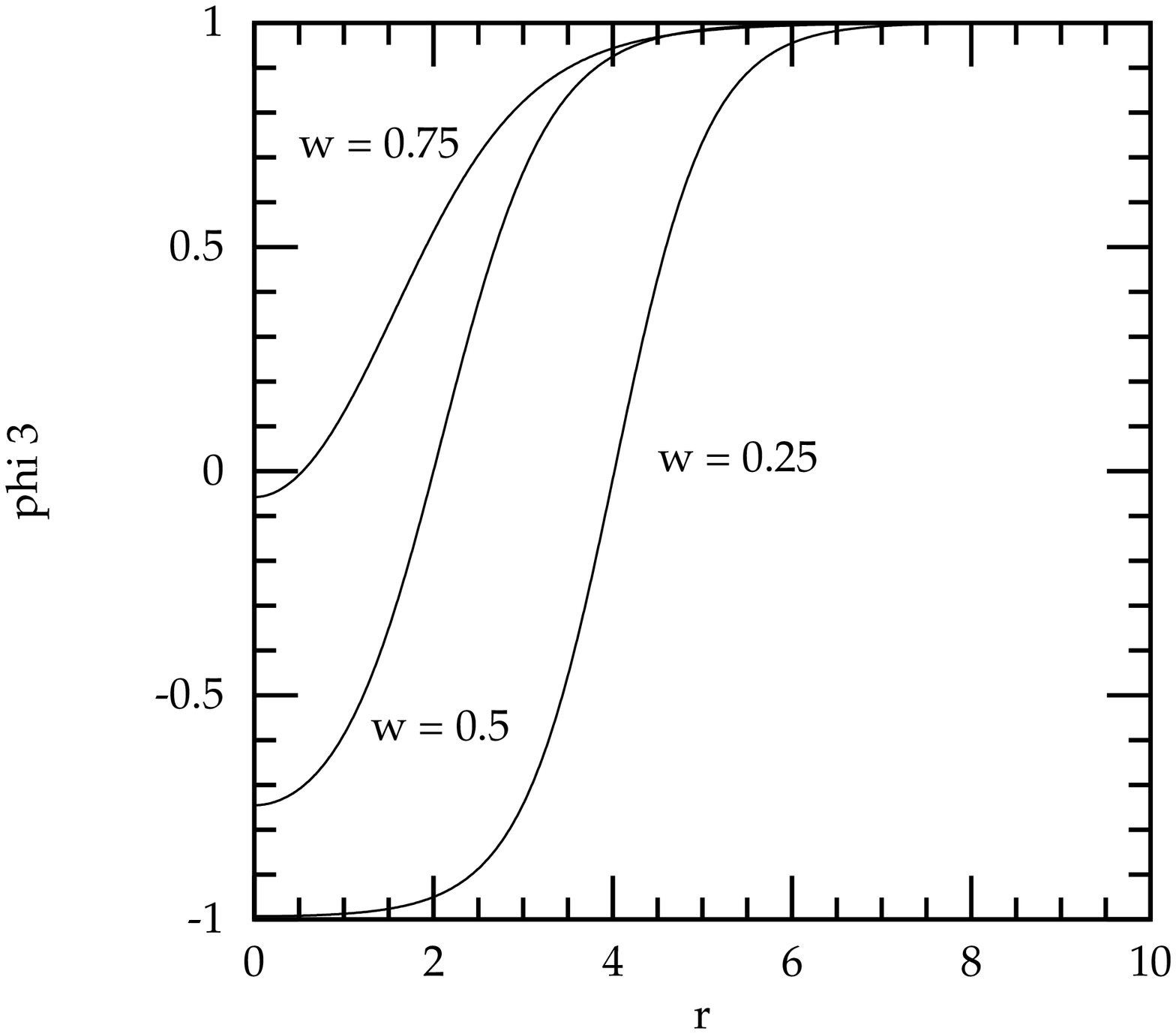}{Non topological case.\break Profile of $\phi_3$ for $\omega$ = 0.25, 0.5 and 0.75}{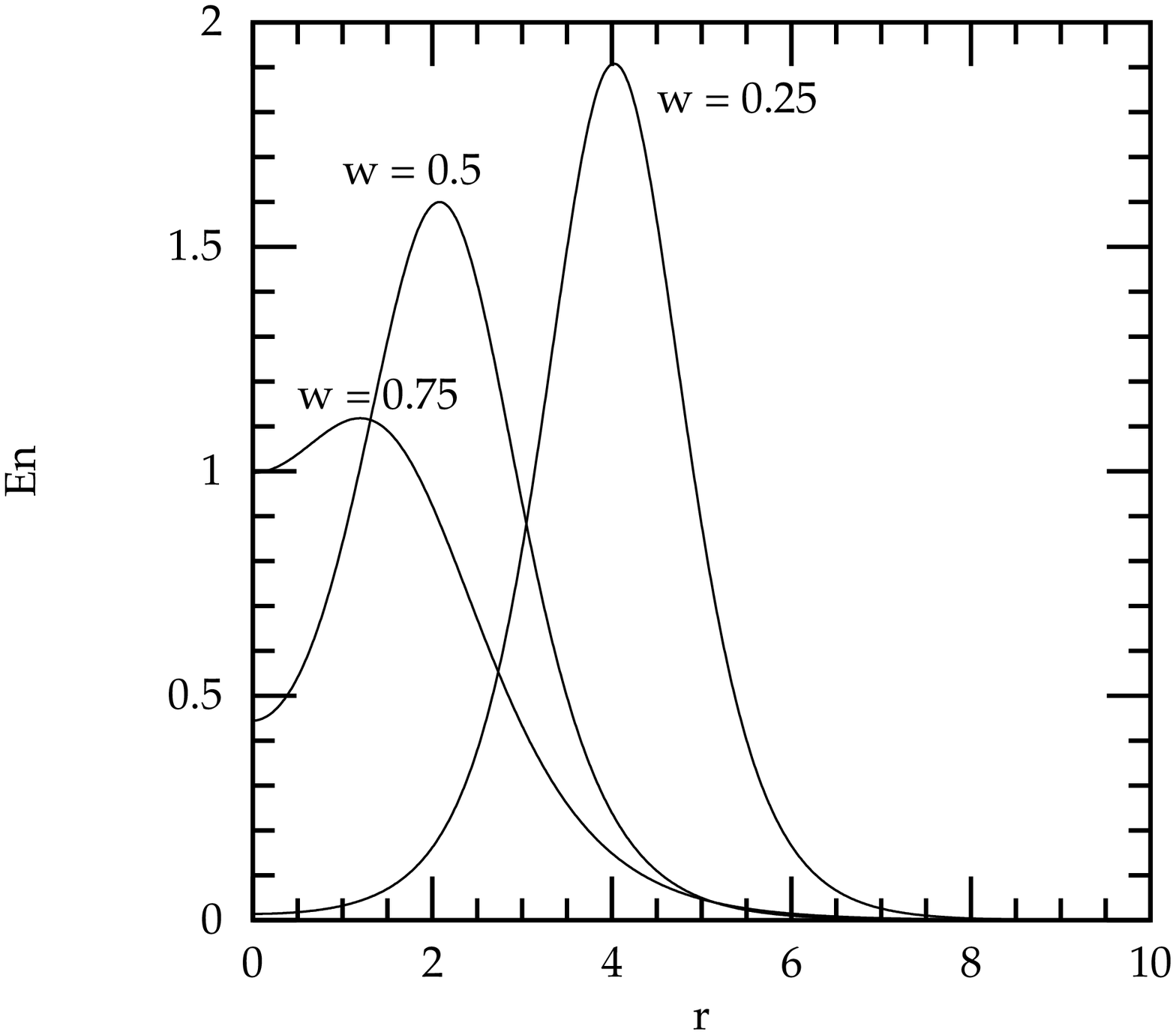}{Non topological case. Energy profile for $\omega$ = 0.25, 0.5 and 0.75}

It is clear from \eEnRealW\ and \eQW\ that the topological density 
of the stationary solutions we have constructed is identically zero. These
solutions are thus of a non topological nature.

To ensure that the solutions we have derived have solitonic 
properties we must verify
 that they are stable. Indeed, it could happen that a radially symmetric
solution is an unstable configuration which desintegrates when it is slightly
perturbed. To analyze this stability numerically we have decided to solve the 
full equation of motion \eEqW\ numerically and we have also looked at the 1+1 
dimensional PDE obtained by
taking the ansatz $w = g(r,t)$, where $g$ is a complex function:
$$
i g_t + g_{rr} + {g_r \over r} - {2 g g_r^2 \over 1+\cmod{g}} 
- A g {1 - \cmod{g} \over 1 + \cmod{g}} = 0.
\eqn\eEqg
$$
This ansatz is still radially symmetric but no specific time dependence is
enforced. The reason for studying \eEqg\ is that the time evolution of its
field progresses faster, 
and as a result, we are able to use finer grids and run our simulation 
for much longer lengths of equation time. 
Both sets of simulations have shown that, indeed, the
solutions we have derived are stable. When perturbed, the ring ``wobbles"
around its position and emits small rings of energy which propagate out
to infinity. By doing so, the ring loses some energy and eventually
settles down to the original unperturbed stationary field configuration.

We have also subjected our solutions to several genuine 2d perturbations.
These studies involved taking one of our configuration and adding to it one 
or two gaussian lumps centered at various points in $x,y$ plane.

In particular we placed our nontopological structure at the origin
and took $\omega=0.5$. Thus, as can be seen from fig 1b, the energy
density extends to $r\sim5$. Then we added to 
$w$ in \eEnRealW\ a gaussian perturbation of the form 
$$
w_{pert}(x,y)= A\,exp\Bigl({(x-a_0)\sp2+(y-a_1)\sp2\over B}\Bigr),
\eqn\epert
$$
and varied the values of parameters $A$, $B$, $a_0$ and $a_1$.
All such perturbations gave qualitatively similar results. The perturbation
deformed the field configuration which then evolved by sending out some
energy and settling to a field configuration corresponding \eEnRealW\
to a value of $\omega$ which, in general, was different from the 
initial value. When the perturbation was not central {\it ie} if
$a_0\ne0$ and $a_1\ne0$ then the resultant field configuration
was deformed in a non-symmetric way and hence it moved in the direction 
opposite to the deformation (similar effects
will be discussed in the next section).

 We have also looked at the effect
of two perturbations (both of type \epert) with 
the opposite values of $a_0$ and $a_1$. This time, as expected,
 the field evolved into a stationary configuration which, incidentally, 
corresponded to $\omega=0.38$. The evolution
was very rapid at first - the energy decreased from 3.6485 at $t=0$
to 2.85 at $t=40$ but then slowed down
so that it reached 2.68 at $t$=615.  All together our
simulations  have demonstrated
the strong stability (and so indestructability) of our non-topological 
solutions. Moreover, in none of the simulations we have performed, have we seen 
the non-topological solutions being destroyed. The only exception to this
(as we will describe later) is when two such configurations merge into one.

\chapter{Moving Non topological Structures}
The periodic configurations \eEnRealW\ satisfying \eEqf have also 
the remarkable property that they
can be made to move at a constant speed if they are deformed in an
appropriate way. If we take as the initial condition for \eEqW 
$$
w = f(r) e^{i (a x + b y) \pi}
$$
where $f$ satisfies \eEqf\ for a certain value of $\omega$ 
and where $k= (a^2+b^2)^{1/2}$ is a constant then the field
configuration describes an extended structure which moves at 
a constant speed in the direction of $(a , b)$. The field still spins at the
frequency $\omega$ but is slightly out of phase:
the spinning at the front is retarded while that of the tail is advanced
relative to the bulk of the structure.

The energy density of such a moving nontopological structure forms 
a ring which is virtually indistinguishable from that of the undeformed 
configuration (\ie correspon-ding to $k=0$). What is more interesting, 
is the topological 
charge density: though the total topological charge is zero, the topological
charge density exhibits a peak and an anti-peak sitting side by side (Fig 2). 
The structure moves in the direction
perpendicular to the line joining the peak and anti-peak,  
leaving the peak on its right-hand side.

\OneFig{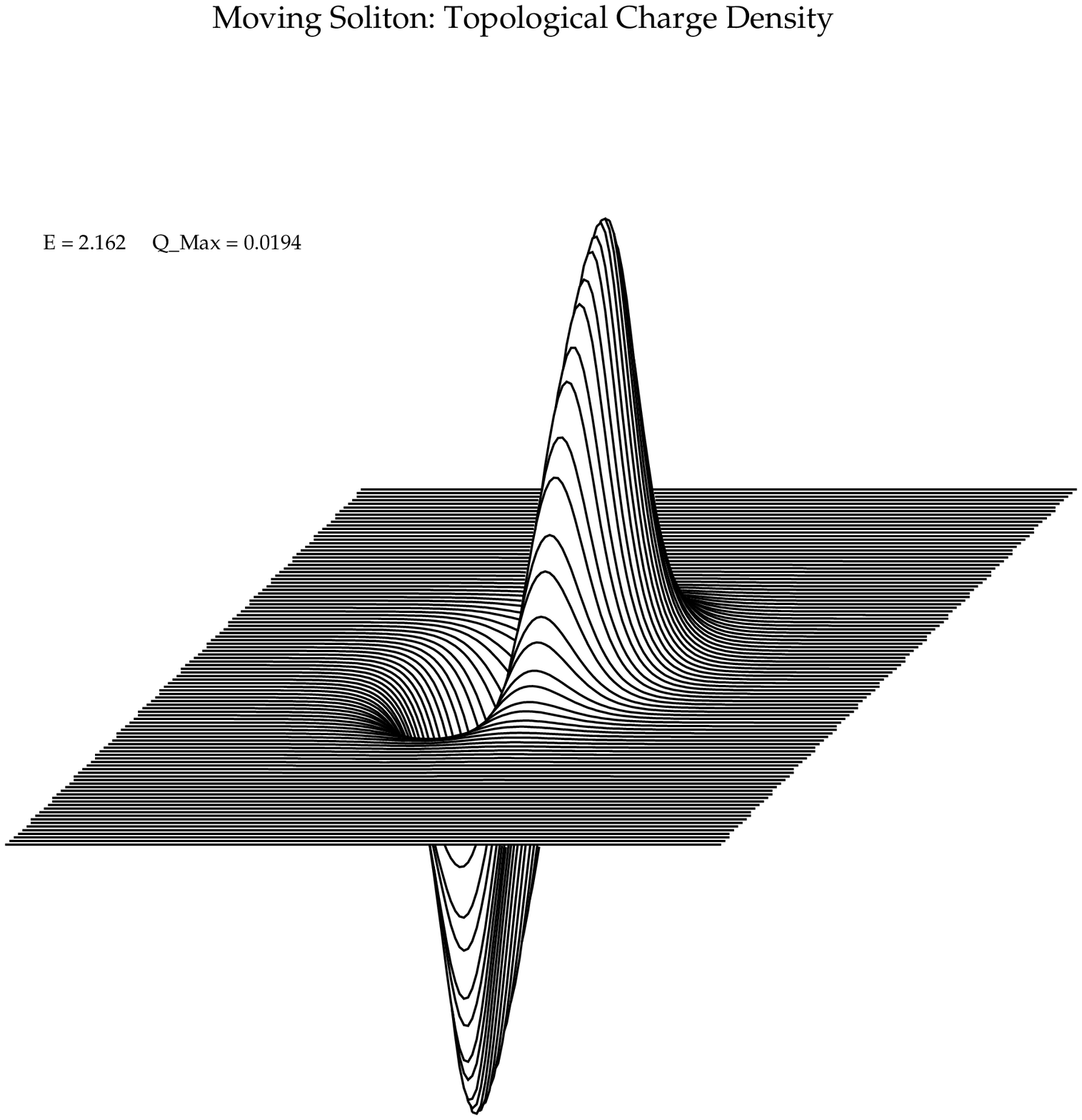}{Topological charge density of a moving structure. 
$(a,b) = (0 , 0.1)$. En = 2.162, Qmax = 0.0194}
 
In Table \NextTableNo\ we present the
speed of translation for a few values of $k = (a^2+b^2)^{1/2}$.

\Table{\Speed}{Speed of deformed structures, $\omega = 0.5$.}{
\centerline{\vbox{\offinterlineskip\tabskip=0pt
\halign{\strut\ #\hfill&\qquad#\hfill&\qquad#\hfill&\qquad#\hfill
&\qquad#\hfill&\qquad#\hfill\cr
k  & 0.025 & 0.05 & 0.1 & 0.15 & 0.2\cr
\noalign{\hrule}
&&&&&\cr
v  & 0.067 & 0.13 & 0.24 & 0.3 & 0.35 \cr
\noalign{\hrule}
&&&&&\cr
E  & 2.04 & 2.06 & 2.15 & 2.3 & 2.5 \cr
}}}
}

As we have moving structures we can study their scattering properties.
To begin with we have scattered 2 structures with $\omega = 0.5$, 
initialy separated by 14 units. They have been made to move  
head-on using the initial condition
$$
w_0 = f(r_1)e^{i \pi (y-b) k} + f(r_2)e^{-i \pi (y+b)k}, 
$$
where $b = 7$, $r_1 = (x^2 + (y-b)^2)^{1/2}$, $r_2 = (x^2 + (y+b)^2)^{1/2}$
and $k = 0.1$. The two structures start by moving towards each other at a 
constant speed.
Then they overlap and eventually perform a 90 degrees scattering.
In Fig 3, we plot the energy density and the topological density of the 
initial configuration, and of the configuration after 40 units of time.
We see clearly that each peak combines itself with the anti-peak of the
other structure to form a new structure that emerges at 90 degrees.

Finally, it would be interesting to know how the non-topological structures 
interact when they are placed at rest next to each other.
To answer this question, we have put two structures (both with $\omega=0.5$) 
at rest so that they overlap a little. We took
$$
w_0 = f(r_1) + f(r_2)
$$
as our initial condition and have observed that the 2 structures initially 
moved towards each other.
Then they scattered at 90 degrees without ever really escaping from each other.
Then they moved backwards and scattered at 90 degrees again, radiating some 
energy. This cyclic motion was repeated many times, and at each step,
the system lost a bit of energy. Eventually, the 2 structures merged into a 
single one which then was spinning at the frequency $0.37$ and had an
energy equal to $2.71$. 

\FourFigsAD{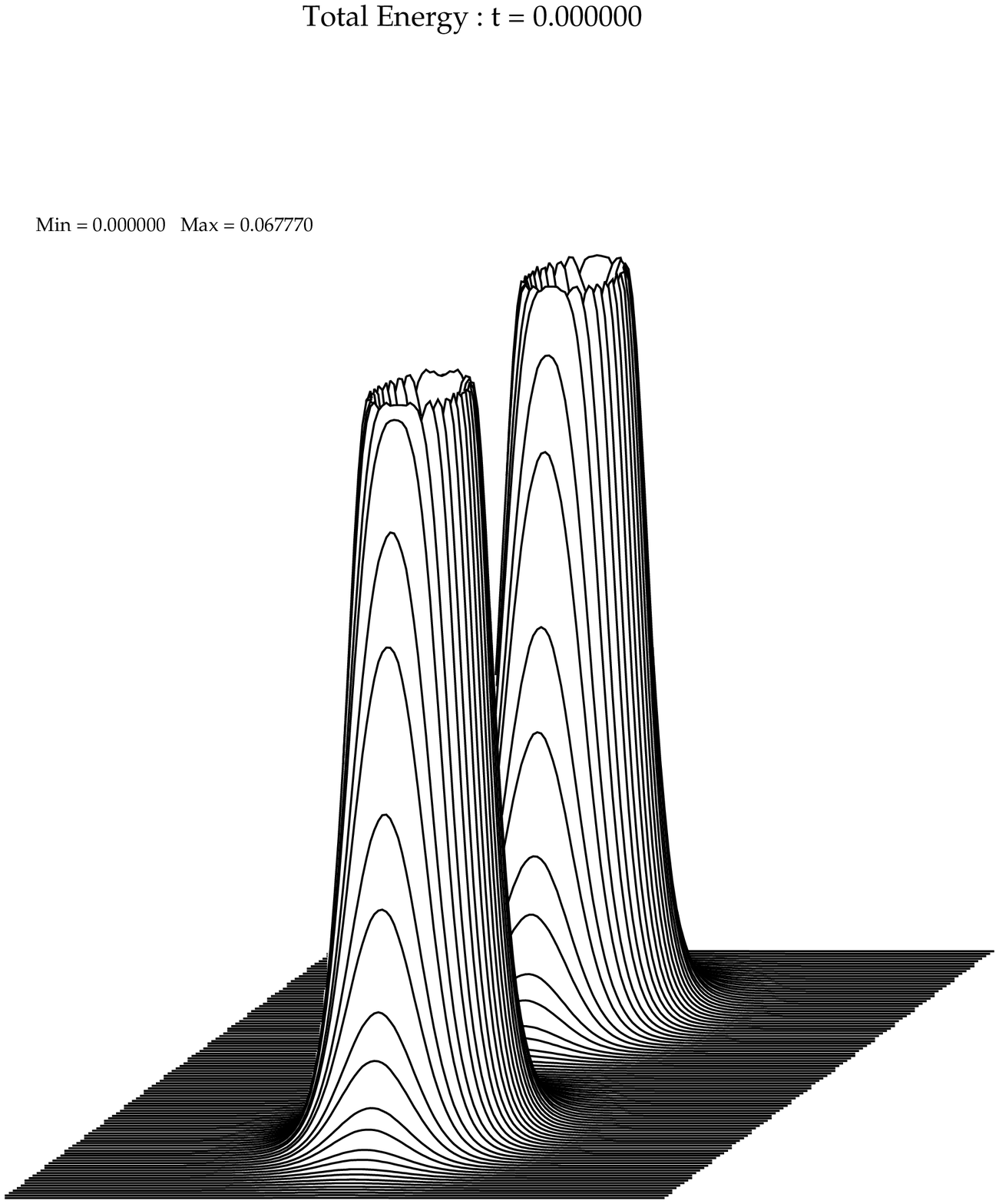}{The periodic structure scattering, Energy density: t = 0}{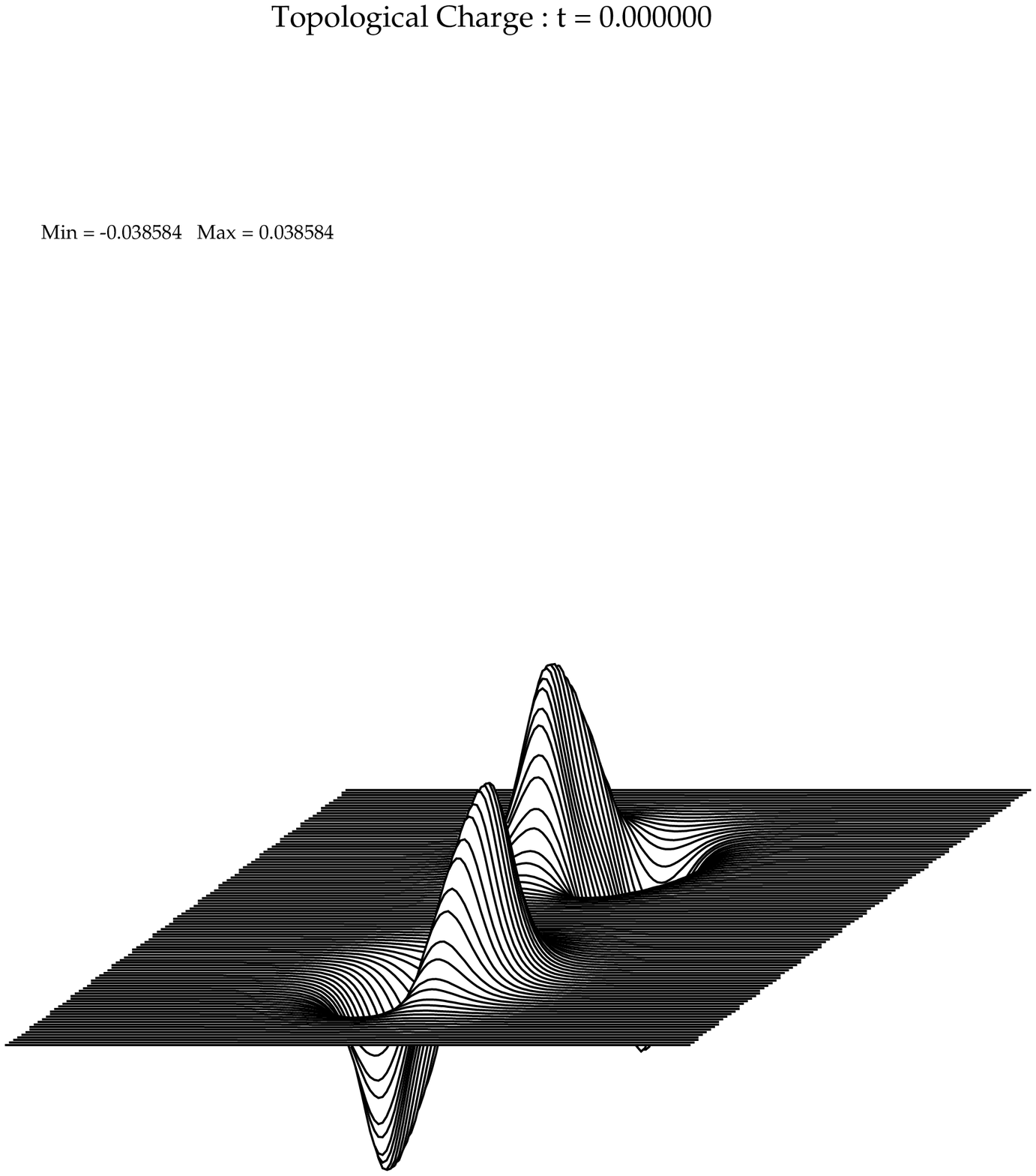}{Topological charge density : t = 0}{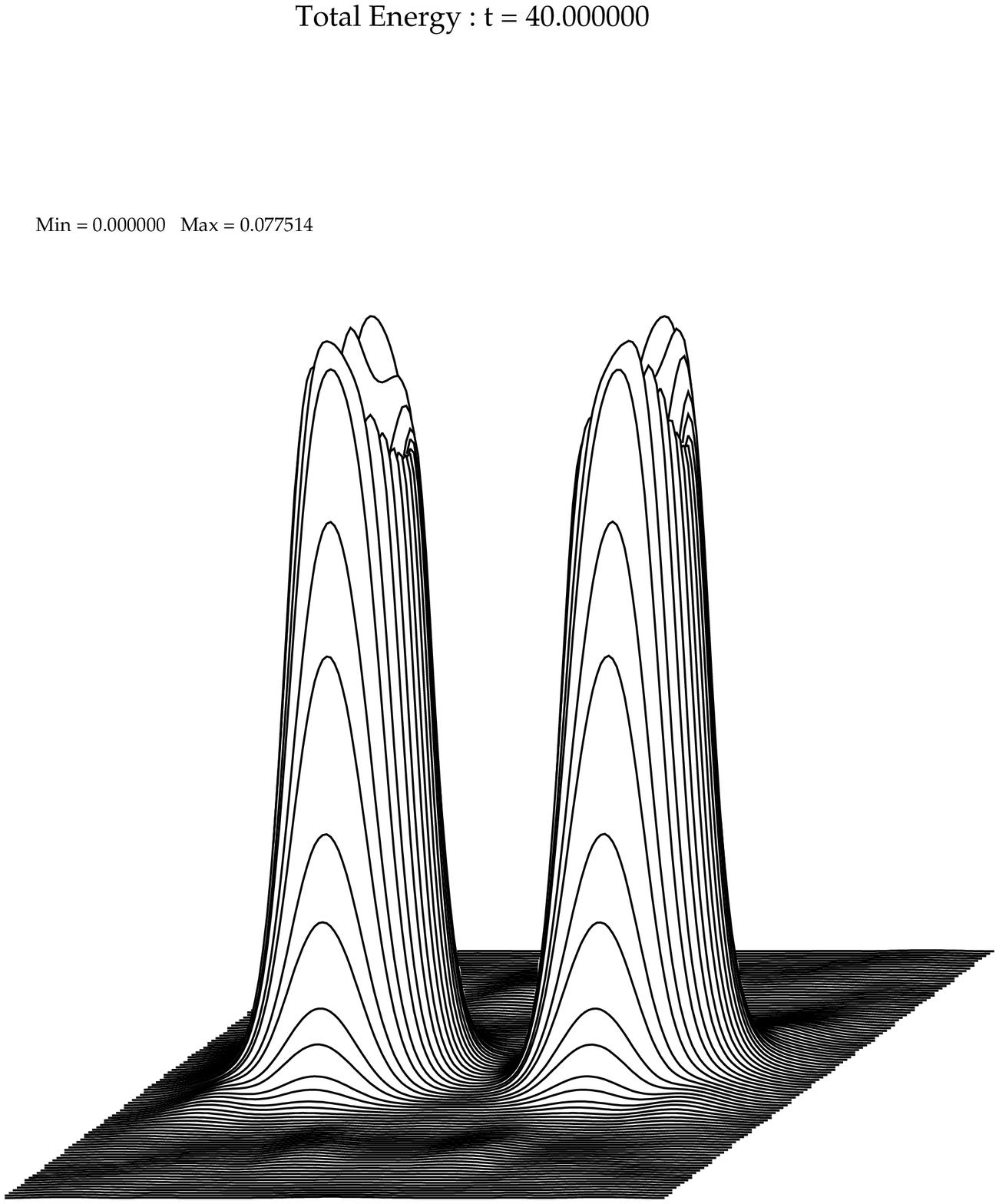}{Energy density after the scattering : t = 40}{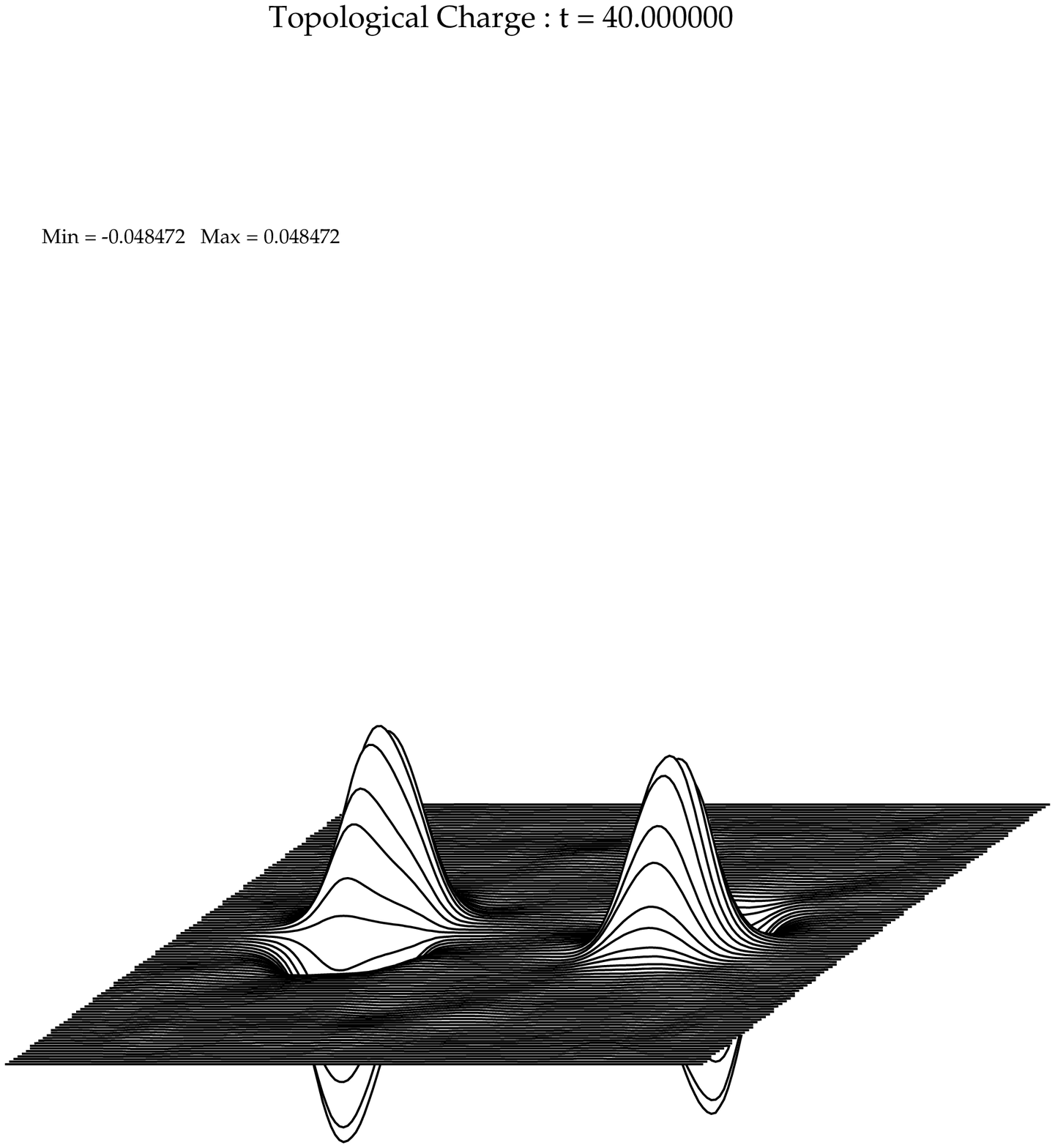}{Topological charge density : t = 40}

We have also looked at the field configurations involving two structures
corresponding to two different values of $\omega$. 
To do this we placed at $x=0$ $y=\pm 4$ the two nontopological field configurations corresponding to $\omega=0.6$ (at $y=4$) and $\omega=0.5$
(at $y=-4$). Recall that as $E\sim {A/\omega}$ the field at $y=-4$
had higher energy. Like in the case mentioned before the two structures
started spinning and attracted each other, performed
a couple of oscillations at $90\sp{\circ}$ and finaly evolved into a single
structure spinning with $\omega\sim 0.42$. Most of the energy decrease
took place during the initial oscillations (in fact by $t\sim 75$ the energy
has already decreased from the initial value of 3.75 to 2.90); the 
single nontopological structure could be easily identified 
by $t\sim 50$ at which time it appeared at $x=0,\;y\sim-1.8$ and then 
proceeded to move slowly towards more negative values of $y$. 
Hence the motion is dominated by the structure of higher energy
and it proceeds in the direction towards its position. It is interesting to 
note that
in both cases the resultant $\omega$ was lower than the initial ones.

This demonstrates that the non-topological nature of our periodic structures 
makes it possible 
for them to merge. As such, the number of structures is not conserved,
but this is not surprising as the structures are nontopological
and so their number is not expected to be conserved.  On the
other hand they are very stable as they don't decay or destroy themselves spontaneously.

\chapter{Topological Periodic Solutions}
In the previous sections we have seen that the easy axis model possesses non 
topological periodic solitons. 
Is it possible to find topological structures 
of a similar nature? We can, for example, take the ansatz
$$
w = e^{-i \omega t} g(r) e^{i n \theta}
\eqn\eEnRealgW
$$ 
where $\theta$ and $r$ are the polar coordinates, $n$ is an integer
and $g(r)$ is a real function. The total topological charge \eQW\ of such a 
configuration is $n$.
Inserting the ansatz \eEnRealgW\ into \eEqW\ leads to the following 
equation for $g$ :
$$
g_{rr} + {g_r\over r} - {2 g g_r^2 \over 1 + g^2}- 
 \Bigl(A  + {n^2 \over r^2}\Bigr) g { 1 - g^2 \over 1 + g^2} + \omega g = 0.
\eqn\eEqg
$$ 
Before we proceed to solve this equation, let us notice that it can be 
derived by minimizing the functional
$$
D = \int r dr \Bigl[ {g_r^2 + {n^2 g^2 \over r^2}+ A g^2\over (1 + g^2)^2} -
 { \omega g^2 \over 1 + g^2} \Bigr].
\eqn\eEng 
$$
We then observe that the change of 
variable $r \rightarrow \lambda r$ leaves the first two terms in $D$ 
unchanged but multiplies the last two by $\lambda^2$. This means that 
the last two terms must add up to zero, and again, we see that
$$
A > \omega.
\eqn\eOmegAbound
$$ 
To integrate \eEqg\ we start by analyzing the asymptotic behavior of
$g$. At infinity $g$ must vanish to ensure that the total energy
\eEnW\ is finite. At the origin, $g$ must be of the form 
$g = r^n(K + Br^2 +O(r^3))$ 
where $K$ is a constant depending on $A$ and $\omega$,
and where $B = -K {A + \omega \over 4 (1+n)}$.
As for the non-topological configurations, we can always scale $r$ to change the
value of $A$. From now on we will thus set $A = 1$. 
\eEqg\ can then be integrated numerically and in Figure 4 we give the
profile $\phi_3$ and the energy density for different values of $\omega$.
Numerically, we find that for $n=1$ the total energy is related to $\omega$ as
: $E = -0.34 + 1.08/\omega$.

The first thing we must check is the stability of such solutions. To do this, 
we have solved \eEqW\ numerically using for the initial condition
\eEnRealgW\ where $g$ satisfies \hbox{\eEqg.} We 
have found that even when we add 
a perturbation to the initial condition the solution is stable. 

In Figure 4 we present the profile and the energy density of the topological
configurations for different values of $\omega$.
\TwoFigsAB{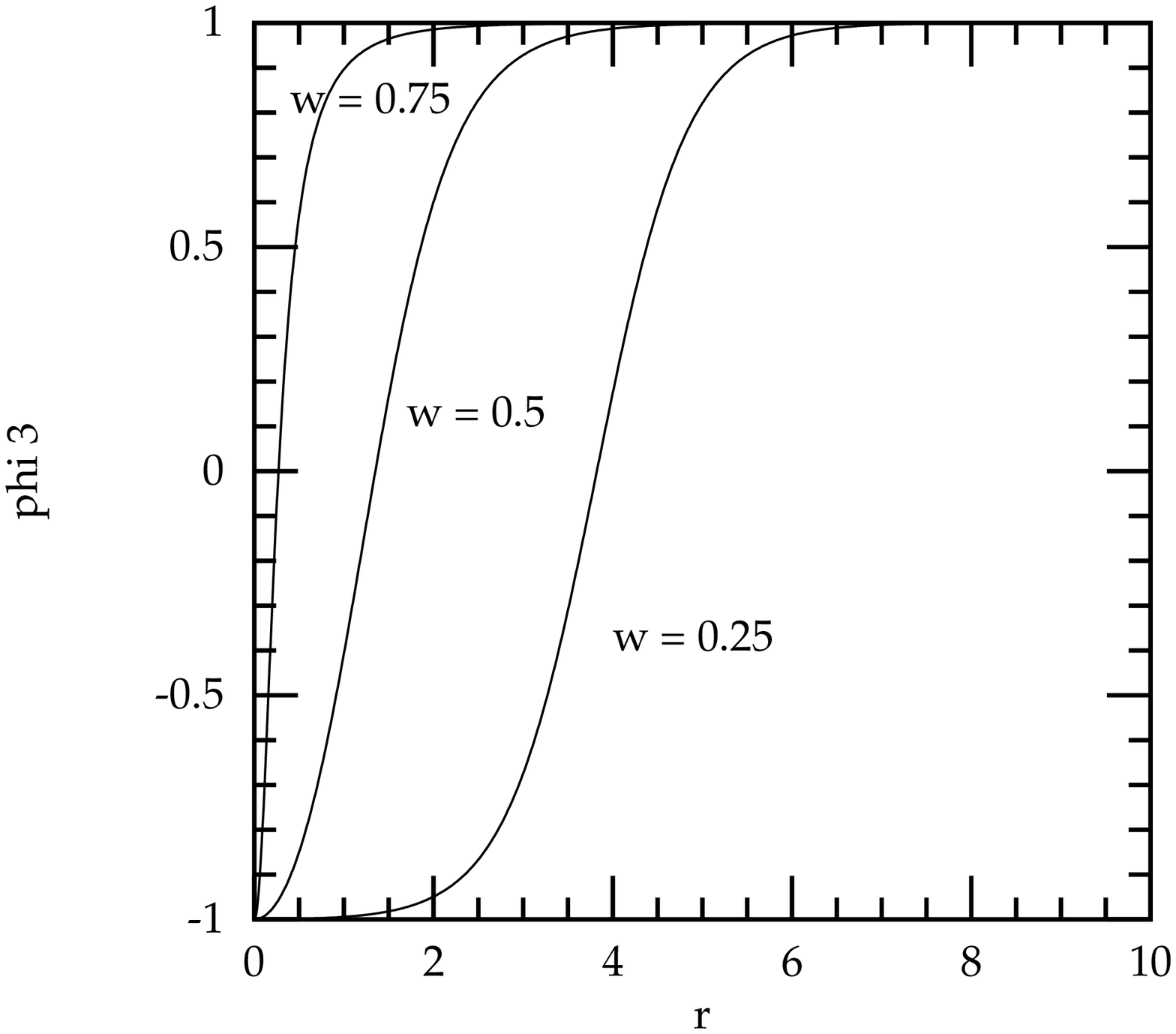}{Topological structure ($n$ $=1$). $\phi_3$ profile for $\omega = 0.25, 0.5$ and $0.75$}{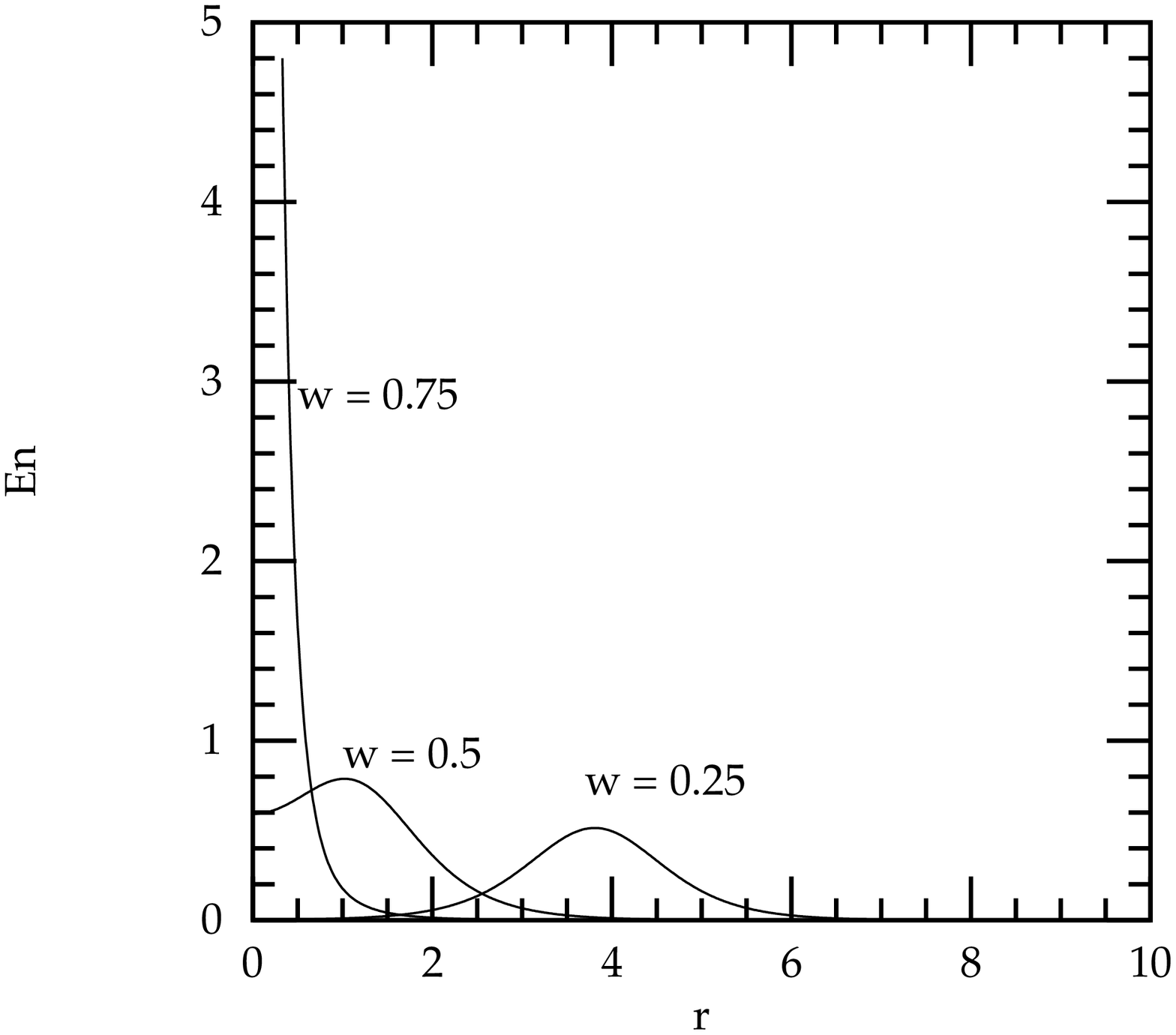}{Energy profile for $\omega = 0.25, 0.5$ and $0.75$ ($\omega = 0.75 : E(0) = 26.53$)}

Finally, let us emphasise the fact that for any value of $N$, \eOmegAbound\ 
gives an upper bound for the largest value of $\omega$ for which a solution
exists.
We have tried to numerically determine 
the largest value of $\omega$ for with there is a solution. 
For $N=0$ we have found solutions for values of omega very closed to $1$. 
The largest value we have had success with was $\omega = 0.99.$ 
For $N$ = 1 we have found solutions for all values of $\omega$ up 
to $\omega=0.95$. 
The most interesting case is $N = 2$: we found that $\omega < 5$. 
The largest successful value for which have found a solution was $\omega= 0.495$
with a total energy equal to $2.0005$.

In Table \NextTableNo\ we give the total energy for some periodic 
solutions for $N = 0, 1$ and $2$.    

\Table{\SolEnergy}{Total energy of periodic solutions ($A=1$)}{
\centerline{\vbox{\offinterlineskip\tabskip=0pt
\halign{\strut\ #\hfill&\quad#\hfill&\quad#\hfill&\quad#\hfill
&\quad#\hfill&\quad#\hfill&\quad#\hfill&\quad#\hfill&\quad#\hfill&
\quad#\hfill\cr
$\omega$  & 0.1 & 0.2 & 0.3 & 0.4 & 0.5 & 0.6 & 0.7 & 0.8 & 0.9\cr
\noalign{\hrule}
&&&&&&&&&\cr
$E (N = 0)$ & 10 & 5.01 & 3.36 & 2.54 & 2.03 & 1.69 & 1.43 & 1.23 & 1.07 \cr
&&&&&&&&&\cr
$E (N = 1)$ & 10.03 & 4.97 & 3.28 & 2.38 & 1.82 & 1.44 & 1.19 & 1.13 & 1.0006 \cr
&&&&&&&&&\cr
$E (N = 2)$ & 7.81 & 3.76 & 2.73 & 2.30 & - & - & - & - & - \cr
}}}
}

\chapter{Interaction between Topological Structures}
Periodic topological structures are genuinely different from non-topological 
ones. The
first evidence of this fact is obtained when two structures are
placed next to each other. For this we define $z_{\pm a} = x \pm a + i y$
and $r_{\pm a} = \mod{z_{\pm a}}$,
and use the following expression as an initial condition
$$
w_0 = g_1(r_a) {z_a\over r_a} e^{i \pi C} + g_2(r_{-a}) {z_{-a}\over r_{-a}}
$$
where $g_1$ and $g_2$ both satisfy \eEqg with the angular velocity
respectively equal to $\omega_1$ and $\omega_2$.
The structures are centered around $a$ and $-a$, and 
the term $e^{i \pi C}$ corresponds to the relative phase between them.
We have performed various simulations for different
values of $C$ and have observed no qualitative difference.

We have taken different values for $\omega_1$ and $\omega_2$, namely: 
$(0.5, 0.5), (0.5, 0.4), (0.6, 0.4)$ and $(0.5, 0.45)$.
When the two values of $\omega$ are different the behaviour is very similar:
the configuration's field starts spining around the center of each structure and
the structures begin to rotate along a circle (clockwise) centered at the 
mid point between them. 
At the same time, they interact 
with each other and, as a result, change their shape and their internal 
speed of rotation. The sizes of the two structures oscillate out of phase:
as one them become spikier and spins faster the other one becomes 
broader and spins slower; then the situation is reversed.
The first maxima are relatively high, but then the field configuration settles
down to what looks like a periodic motion. 

One could have expected that the two structures would progressively
tune their angular rotation speed to eventually forming a configuration
with both of them moving along a circle and spinning internally at the same 
speed.
Our simulation shows that this is not the case: in each simulation the final 
configuration correponded to two stuctures moving along a circle and exchanging 
energy all the time. This exchange of energy makes them oscillate in size 
and modulates their internal spinning frequency.

This is partially confirmed by the simulation when the two $\omega$ have 
identical values. 
In this case, as expected, the fields start to spin with the angular frequency 
$\omega$. At the same time the 2 structures start rotating around their 
midpoint. As the evolution progresses, the field configuration in  
the region between the two structures, 
initially of the form $\phi = (0,0,1)$, takes non trivial values in the
$\phi_1, \phi_2$ plane ($\phi_3 = 0$). Apart from a regular rotation around
the $\phi_3$ axis the field is constant in this region. The energy density 
it generates is very small (virtualy invisible in the energy density graphs.) 
However, after a while, the region of constant $\phi$ starts to move between 
the 2 structures along
a trajectory shaped like an ``8'' for which the structures are located
in the middle of the two circles. This time, the size of the two structures 
oscillates in phase, but eventualy (at the fifth maximum in our simulation), 
as the energy is exchanged between the two
structures, one of them becomes much larger than the other one and it spins 
internally
much faster. As the structures becomes very spiky the number of lattice points 
supporting it becomes too small and the numerical simulation 
breaks down. It is difficult to decide, from a numerical point of view,
if this blowup is a genuine effect or simply a numerical artefact, but
we believe it is more likely to be a numerical problem rather than a genuine
blow up. To resolve this problem we would have to take many more points for our
lattice, something we are not able to do with the computers at hand.

Finally we have put 2 stuctures of different but very close angular frequencies
$(0.5, 0.49)$ next to each other. The behaviour was very similar to what we 
saw when the frequencies were the same. Again, we are not
able to determine whether this is a numerical problem or if the blow up 
is genuine.  

Another interesting observation is the interaction between a structure and an 
anti-structure (\ie a configuration with N = -1.) 
We can indeed put the two of them next to each other by using as
the initial condition
$$
w_0 = g(r_a) {z_a\over r_a}  - g(r_{-a}) {z^a_{-a}\over r_{-a}}.
$$
In this case, the structure and the anti-structure move along approximately 
straight lines and the direction of motion is such that when looking forward,
the structure is on the right-hand side. 
The lines are almost straight with an extra periodic oscillation imposed on 
them. In this they resemble the motion seen in [\RNPZ] and 
\REF\RStraTam{\rStraTam}[\RStraTam].

We have performed different simulations for different values of $\omega$
and different value of the separation parameter $a$. 
We have found that the closer the structures are to each other, 
the faster they move. 
In Table \NextTableNo, we give the speed observed in each of the simulation 
we have 
performed.

\Table{\SolASolSpeed}{The speed of Structure anti-Structure pairs.}{
\centerline{\vbox{\offinterlineskip\tabskip=0pt
\halign{\strut\ #\hfill&\qquad#\hfill&\qquad#\hfill&\qquad#\hfill
&\qquad#\hfill&\qquad#\hfill&\qquad#\hfill\cr
$\omega$  & 0.4 & 0.4 & 0.5 & 0.5 & 0.6 & 0.6 \cr
\noalign{\hrule}
&&&&&&\cr
$a$  & 0.2 & 0.3 & 0.2 & 0.3 & 0.2 & 0.3 \cr
\noalign{\hrule}
&&&&&&\cr
v  & 0.18 & 0.14 & 0.21 & 0.12 & 0.11 & 0.018\cr
}}}
}

We have also looked at the interaction when the structures had
different angular frequencies. The main evolution was the same as before,
the pair moved along approximately straight lines at a constant speed, but 
the deformation of the structures was more irregular and the oscillations 
of the structures along the lines where different.

\chapter{Magnetic Bubbles}
We mentioned in the introduction that the solutions of \eEqW\ when there
is no magnetic field, $A=0$, are given by holomorphic functions. The simplest
solution is the so called magnetic bubble, which has a topological charge 
equal to $1$ and is given by $w = \lambda (x + i y)$,
where $\lambda$ is a constant that fixes the size of the bubble. It would be 
interesting to know what happens when we take such a configuration as the
initial condition when $A$ is non-zero. This could be thought of as describing
a system containing an isolated magnetic bubble which is suddenly immersed
in a constant magnetic field.

Looking at \eEqPhi or \eEqW we notice that we have two parameters at hand, 
$A$ and $\lambda$, but we can fix one of them by performing a rescaling of the
coordinates. For our simulations, we have decided to keep $\lambda$
constant and vary $A$; this way we can directly analyse how the strength 
of the magnetic field affects the magnetic bubbles. From now on we will
thus assume that $\lambda=1$.

What we have done, is to take the initial condition $w = x + i y$ and solve
\eEqPhi\ for different values of $A$. When $A=0$ the magnetic bubble is a 
static solution but as soon as there is a magnetic field the field starts to 
spin.
In other words, the field takes the form of 
$w = g(r,t) e^{i (\theta + \omega t)}$
where $r,\theta$ are polar coordinates, $g(r,t)$ is a complex function 
(which varies with time) and $\omega$ is the angular speed of precession of
the field. Notice that the initial condition is simply $t=0$, $g(r) = r$.
We have solved numerically both \eEqPhi\ and \eEqg\ and have found that the 
angular speed does not change with time. 
The main time evolution comes from the profile $g(r,t)$.
The energy density shows that the bubble emits rings of energy which propagate
towards infinity (we thus had to implement some absorption scheme to stop the 
wave from reflecting itself on the grid boundary). 
As the bubble radiates away some energy it
progressively settles to a periodic configuration which is nothing but the
solution described in section 4 with the corresponding value of
$\omega$. 
We have not found an argument to predict the value of $\omega$ 
(\ie how $\omega$ depends on $A$.)
In Table \NextTableNo\ we give some of the values we have observed in our 
simulations.
When reading Table \NextTableNo,
one must remember that the solutions described in section 4 assume that $A=1$.
Thus the angular velocity listed in the table must be divided by the 
corresponding value of $A$ before we can compare these solutions with those 
shown in Fig 4. 

\Table{\HolSol}{Magnetic bubble energy and their angular velocity.}{
\centerline{\vbox{\offinterlineskip\tabskip=0pt
\halign{\strut\ #\hfill&\qquad#\hfill&\qquad#\hfill&\qquad#\hfill
&\qquad#\hfill&\qquad#\hfill&\qquad#\hfill\cr
$A$  & 0.1 & 0.2 & 0.5 & 1 & 2 & 5 \cr
\noalign{\hrule}
&&&&&&\cr
$\omega$  & 0.072  & 0.13  & 0.278 & 0.48 & 0.82 & 1.47 \cr
\noalign{\hrule}
&&&&&&\cr
$E$  & 1.15 & 1.30 & 1.59  & 1.90 & 2.33 & 3.35 \cr
\noalign{\hrule}
&&&&&&\cr
}}}
}

\chapter{Conclusions}
We have shown in this paper that the Landau-Lifshitz equation with an
anisotropic term has various types of stable solutions periodic in time.
All the solutions have the common property that the field spins internally
at a constant speed. The solutions can have any topological charge 
or be topologicaly trivial. All the solutions are stable, in the sense that they
can't be destroyed by small ( and not so small) perturbations.
The only exception are the non-topological solutions
which can merge together to form a new solution of the same type.

The non-topological structures have other surprising properties: they can be 
made to move at a constant speed by being deformed in a specific way. 
Such moving configurations scatter at 90 degrees. 

The topological structures have different properties. When two 
of them are close to each other they move along a circle and exchange
energy. Their size and internal angular speed of precession
are periodic functions of time.
Structures and anti-strcutures do not anihilate but move along essentially
straight lines modulated by small oscillations.

Finally, when taken as an initial condition, holomorphic magnetic bubble
modify themselves to become stationary solutions with internal field 
precession.

\ack
This work was initiated soon after we had some fruitful discussion with
M. Lakshmanan and M. Daniel. We want to thank them both for sharing with us 
their knowledge of the Landau-Lifshitz model and for drawing our attention to 
ref [\RLak]. 

We also wish to thank N. Papanicolaou and G. Stratopolos for helpful 
conversations.

\refout 
 
\end